# Analysis of Learner Independent Variables for Estimating Assessment Items Difficulty Level

Shilpi Banerjee, International Institute of Information Technology, Bangalore, India
shilpibanerjee1@gmail.com, 09731128466
N.J.Rao, International Institute of Information Technology, Bangalore, India
njraoiisc@gmail.com, 09343537734

## ABSTRACT

The quality of assessment determines the quality of learning, and is characterized by validity, reliability and difficulty. Mastery of learning is generally represented by the difficulty levels of assessment items. A very large number of variables are identified in the literature to measure the difficulty level. These variables, which are not completely independent of one another, are categorized into learner dependent, learner independent, generic, non-generic and score based. This research proposes a model for predicting the difficulty level of assessment items in engineering courses using learner independent and generic variables. An ordinal regression model is developed for predicting the difficulty level, and uses six variables including three stimuli variables (item presentation, usage of technical notations and number of resources), two content related variables (number of concepts and procedures) and one task variable (number of conditions). Experimental results from three engineering courses provide around 80% accuracy in classification of items using the proposed model.

## 1. Introduction

Course outcomes represent what the students should be able to do at the end of a course as per the instructor. Course outcomes of an engineering course are designed by the instructor to meet a subset of Program Outcomes identified by the accreditation agency of the country and Program Specific Outcomes chosen by the Department offering the program. Each course outcome statement is mapped to one of the cognitive levels and one or more categories of knowledge of Anderson-Bloom taxonomy (Anderson, Krathwohl, and Bloom, 2001). The assessment items associated with a course outcome are termed to be aligned if they belong predominantly to the same or lower cognitive level as that of the course outcome. Mastery of learning can be defined in terms of difficulty levels of items in assessment instruments.

The way each instructor perceives the difficulty level of an assessment item is not the same, and he/she may associate different combinations of values associated with an assessment item and their values of a perceived difficulty level. An attempt is made in this paper to identify the latent process used by the instructors for predicting the difficulty level as low, moderate or high. Proposed model considers learner independent and generic (subject non-specific) variables, and does not use evaluation scores. Following section describes the existing frameworks for estimating the difficulty.

## 2. Research Background

One of the most widely used estimates of item difficulty level is the proportion of students who correctly answer an item (Liu et al., 2010). Difficulty is estimated, in this approach, using evaluation scores. In another scenario, it is defined as the amount of effort needed to answer an item. Inherent difficulty of an item is estimated, in this approach, by observing item variables. Moreover, the meaning and interpretation of difficulty varies with the application for which it is estimated. Approaches to estimation of difficulty level are proposed in literature in the contexts of intelligent tutoring systems (ITS), computer adaptive tests (CAT) and class room assessments. Difficulty is estimated dynamically for ITS by considering the performance of individual students in an item (Jiu and Park, 2004; Hatzilygeroudis et al., 2006; Li and Sambasivam, 2003). Difficulty level of an item is adjusted in CAT to the students ability level (Karahoca, Karahoca, and Ince, 2009; Kunichika et al,. 2002). Affective parameters play an important role for estimating difficulty in ITS and CAT. Static measures of difficulty of items are proposed for classroom assessments by considering content, stimulus, response and task variables. Several approaches are used to measure difficulty in each of these applications. Rasch models, Item Response Theory and Linear Logistic Test Model were used for CAT. Knowledge map, concept map, ontology and artificial intelligence techniques were used for ITS. Psychological and statistical models were used for classroom assessments (Banerjee, Rao, and Ramanathan, 2015). *This paper proposes a method of estimating the difficulty of formative and summative assessments.* The variables used for existing approaches to estimate item difficulty level are either learner dependent or independent. There can be multiple reasons, other than inherent difficulty of an item, because of which a student finds an item more difficult which includes task motivation, anxiety, openness to experience, willingness to communicate (Gan, 2011), working memory, task switching, aptitude, self-efficacy, openness and implicitness (Hughes, Pollitt, and Ahmed, 1998).Thus, an assessment item is perceived differently by different students with regard to the difficulty level. The measures of the learner dependent variables for one batch of students may not be applicable to another batch of students (Dhillon, 2003). *The proposed approach uses only the learner independent variables for estimating difficulty level.*

Difficulty can be estimated using evaluation scores or by using item variables. True score for a student may be different from the evaluation score due to the presence of random and systematic factors encountered in the process of administering assessment and evaluation. Random factors are entirely due to chance and are due to learner dependent and assessment environment specific variables. Assessment environment specific variables include noise distractions, poor lighting and uncomfortable room temperature. Systematic factors include inter-rater and intra-rater factors (Moskal, 2000) which occur when two raters rate the same item differently or when one rater rates same item for two different students differently. Therefore, difficulty estimate based on evaluation scores may not give an actual estimate of item difficulty. There are other difficulty estimation frameworks which take into account opinions of subject matter experts (SME). *The item difficulty level is estimated before offering it to students by considering only the item variables in the proposed approach.*

Another categorization of difficulty estimation approaches is based on generic and non-generic variables. Generic frameworks are not specific to any particular subject while non-generic frameworks are applicable for all subjects. Every subject has some characteristics that make it different from other subjects. Also, items of the same difficulty level for two different subjects may not be comparable. Non-generic variables used in existing approaches for courses on social sciences and humanities includes word count, prepositional phrases, modal verbs, negations, grammatical structure and syntactic structure (Aryadoust, 2013). *The proposed approach to the estimation of difficulty level of assessment items in engineering courses is based only on generic variables.*

The present study aims to gain a greater understanding of variables that are likely to influence the difficulty level of assessment items in engineering courses. Learner independent and generic variables that characterize the difficulty level are selected from the literature. Similar variables were grouped followed by a statistical analysis of the grouped variables. This study addresses the following research objectives:

1. *Identify the learner independent and generic variables used for estimating difficulty level of assessment items.*
2. *Propose a model for predicting item difficulty level with minimum number of variables and with maximum accuracy.*

3. Learner Independent and Generic Variables

Difficulty corresponding to an item resides in the associated task, content, stimulus and psychological construct (Osterlind and Friedman, 1999). Task difficulty refers to the difficulty that the students face when they generate their responses. Content difficulty is related to the elements of knowledge including facts, concepts and procedures. Stimulus difficulty is related to the manner in which the item is presented to the students which includes words, phrases and information which is packed along with the item (Cheng, 2006). Psychological construct deals with learner dependent variables. Task, content and stimulus variables are identified for estimating assessment item difficulty level by grouping similar variables based on their usage and implementation. A group of expert instructors from engineering institutes were consulted on the validity of the variables selected by the researchers. Coding rules were proposed for the chosen variables which primarily focus on estimating the difficulty level of an item and not estimating the absolute value of difficulty. Therefore, detailed scale is avoided which might not be necessary in this case.

Abstraction refers to the extent to which the student has to deal with ideas rather than concrete objects to answer an item (Ahmed and Pollitt, 1999). This parameter is very essential and heavily contributes to item difficulty. It is considered in the proposed difficulty estimation approach in the form of number of inferences, number of assumptions and item presentation.

## 3.1 Variables not considered in the Proposed Approach

*(1) Hierarchy of Cognitive Processes:*

Because of the hierarchical nature of cognitive levels, there is a notion among some instructors and researchers that items measuring abilities of student for higher cognitive processes are more difficult than items measuring activities lower cognitive processes (Pollitt and Ahmed, 1999; Matters, 2009; Liu et al., 2010). An attempt to resolve the fuzziness between the terms difficulty hierarchy and cognitive process hierarchy is made by conducting an experiment for the hypothesis, Assessment items at higher cognitive level are not necessarily of higher difficulty levels than assessment items at lower cognitive levels (Banerjee, Rao, and Ramanathan, 2015). It was found out that cognitive level and difficulty level are not necessarily related. Therefore, hierarchy of cognitive processes is not considered for the proposed approach.

*(2) Amount of time:*

It indicates total time that student has spent to answer an item correctly (Koutsojannis et al., 2007). Difficulty is not determined by the amount of time taken for responding, as more marks are assigned to the item which requires more time to respond.

*(3) Type of constraint:*

Students generate a potential solution and then test it against the constraints in the item. Item difficulty increases with the nature and type of constraints (Katz, Lipps, and Trafton, 2002), which are subject-specific. Proposed approach of estimating difficulty uses only the generic variables, the type of constraint is not considered in the proposed approach.

*(4) Number of prerequisite concepts:*

Difficulty increases as number of pre-requisite concepts increases (Liu et al., 2010). When a student wants to master a unit of study, they must completely grasp all the related prerequisite conceptual knowledge. The difficulty of an item is not considered to be dependent on the number of prerequisite concepts as all learners are assumed to have an understanding of the prerequisite concepts.

## 3.2 Variables Affecting Assessment Item Difficulty

From the empirical findings, following fifteen hypotheses were derived in response to research objective 1. Number of unknowns ($T_1$), number of conditions ($T_2$) and numerical complexity ($T_3$) were identified as task difficulty variables. Number of facts ($C_1$), number of concepts ($C_2$), number of procedures ($C_3$), combination of knowledge elements ($C_4$) and number of prerequisite course outcomes from the same course ($C_5$) were identified as content difficulty variables. Item presentation ($S_1$), number of hints ($S_2$), independency of unknowns ($S_3$), usage of technical notations ($S_4$), number of inferences ($S_5$), number of resources ($S_6$) and number of assumptions ($S_7$) were identified as stimuli difficulty variables.

*H1: Assessment item difficulty level increases as the number of unknowns ($T_1$) in an item increases.*

An item has some unknown attributes and some given attributes. As $T_1$ increases, the number of facts to be recalled, concepts to be understood and procedures to be mastered also increases. Higher $T_1$ leads to increased processing load on students (Kuo et al., 2004).

*H2: Assessment item difficulty level increases as the number of conditions ($T_2$) in an item increases.*

Conditions represent the process(s) the student is expected to follow in responding to an item. An item becomes more difficult as more conditions have to be satisfied to reach a potential solution. More conditions associated with an item results in a more difficult item (Li and Sambasivam, 2003; Fisher-Hoch and Hughes, 1996; Lumley et al., 2012; Katz, Lipps, and Trafton, 2002).

*H3: Assessment item difficulty level increases as the item has more numerical complexity ($T_3$).*

An intuitive expression of $T_3$ would be the larger a number in an item is, the more difficult it should be, since it is harder to do calculations involving larger numbers (Kuo et al., 2004; Lee and Heyworth, 2000). The probability of errors associated with solving an item with larger numbers is also more.

*H4: Assessment item difficulty level increases as the number of facts ($C_1$) to be recalled for attempting an item increases.*

Facts consist of the basic components student should be aware for becoming familiar with the course or attempting an item in it. Facts include definitions, equations, formulae, relations and some specific quantitative data. Difficulty increases as $C_1$ associated with an item increases, as student needs to recall those facts for attempting an item (Embretson and Daniel, 2008).

*H5: Assessment item difficulty level increases as the number of concepts ($C_2$) for attempting an item increases.*

Concepts include knowledge of categories and classifications and the relationships between and among them. Difficulty is a function of total number of related concepts with an item (Hatzilygeroudis et al., 2006; Guenel and Asliyan, 2009; Koutsojannis et al., 2007; Cheng, 2006). If too many concepts are associated with the item presented, the risk of error increases (Klemola, 2000).

*H6: Assessment item difficulty level increases as the number of procedures ($C_3$) to be mastered associated with an item increases.*

Procedures are operations that students are required to perform on input data for executing a task. Generally the more procedures required in a calculation, the more chance there is to make an arithmetic error as large number of procedures over-loads working memory and information is likely to be lost (Fisher-Hoch and Hughes, 1996; Fisher-Hoch, Hughes, and Bramley, 1994). Assessment items that require more steps in a solution are more difficult than items that require fewer steps (Ahmed and Pollitt, 1999; Lee and Heyworth, 2000; Embretson and Daniel, 2008; Cheng 2006; Ogomaka, Nosike, and Akukwe, 2013), as the cognitive demands the task makes and the amount of information the student is expected to process increases with more number of steps (Gan, 2011).

*H7: Assessment item difficulty level increases as it deals with combination of knowledge elements ($C_4$).*

Difficulty increases as the number of knowledge elements for an item increases. Assessment items that assess students on two or more knowledge elements are generally more difficult than assessment items that assess them on a single knowledge element (Pollitt and Ahmed, 1999; Ahmed and Pollitt, 1999; Fisher-Hoch, Hughes, and Bramley, 1994; Cheng, 2006). Four general types of knowledge (Anderson, Krathwohl, and Bloom, 2001) proposed relevant across all disciplines include Factual, Conceptual, Procedural and Metacognitive knowledge. The categories of knowledge specific to Engineering (Vincenti, 1990) in addition to the four general categories, are Fundamental Design Concepts, Criteria and Specifications, Practical Constraints and Design Instrumentalities. Each assessment item consists of a general/ engineering specific knowledge element or a set of these knowledge elements. Proposed approach considers only the general types of knowledge elements.

*H8: Assessment item difficulty level increases as the number of prerequisite course outcomes from the same course ($C_5$) for an item increases.*

Every engineering course has 6 to 8 course outcomes. Assessments help to evaluate if students have acquired these course outcomes well. Often, the earlier course outcomes are prerequisites for latter ones. Therefore, an assessment item belonging to latter course outcomes is more difficult as compared to assessment item belonging to earlier ones as student needs to achieve mastery in earlier course outcomes to do well in latter course outcomes (Klemola, 2000; Liu et al., 2010; Khan, Hardas, and Ma, 2005).

*H9: Assessment item difficulty level is more for complex item presentation ($S_1$).*

The manner in which information is packed in an assessment item affects the difficulty level of the item (Perkins, Gupta, and Tammana, 1995; Chon and Shin, 2010). Same item can be presented to students in many different ways. For some items, irrelevant information is purposefully introduced which can distract students or in some cases insufficient information is given and student needs to draw a lot of deductions (Fisher-Hoch, Hughes, and Bramley, 1994). Item appears more difficult if students are unfamiliar with the context which includes infrequent words and unfamiliar topic (Sung, Lin, and Hung, 2015).

*H10: Assessment item difficulty level decreases as the number of hints ($S_2$) in an item increases.*

If the concept or procedure is given in an item, student will not have to deduce the topic to which the question is related to. Student finds easy to attempt an item if prompts are given for forming a strategy to approach an item (Fisher-Hoch and Hughes, 1996; Fisher-Hoch, Hughes, and Bramley, 1994; Chon and Shin, 2010; Schmeiser and Whitney, 1973).

*H11: Assessment item difficulty level decreases if the unknowns in an item are independent ($S_3$) of each other.*

An assessment item is said to be interlinked if the unknowns are defined relative to each other (Ahmed and Pollitt 1999; Fisher-Hoch, Hughes, and Bramley, 1994). Relative definitions of unknowns determine item difficulty (Embretson and Daniel, 2008). An item in which a unistructural response is expected is generally easier than an assessment item in which relational response is expected (Gan ,2011).

*H12: Assessment item difficulty level increases if technical notations ($S_4$) are not used in an item.*

Often, technical language is used in engineering assessment items where complex mathematical representation are present as part of the item. Much of the formalism is already been done for these items. Such items are easier to interpret as compared to items for which mathematical formalism is expected from students (Fisher-Hoch, Hughes, and Bramley, 1994; Turner and Adams, 2012).

*H13: Assessment item difficulty level increases as the number of inferences ($S_5$) in an item increases.*

Inferences related to an assessment item are more if it deals with ideas rather than concrete objects or phenomena to answer the item (Ahmed and Pollitt, 1999). The amount of effort in analyzing the item statement increases as $S_5$ increases (Nakamura, Kuwabara, and Takeda, 1998). Difficulty is a function of degree of inferential processing in an item (Ozuru et al., 2008).

*H14: Assessment item difficulty level decreases as the number of resources ($S_6$) in an item increases.*

Resources refer to the diagrams, tables, pictures, graphs or photographs provided with the item. Students need to form a mental representation themselves if the resources are not provided with the item (Ahmed and Pollitt, 1999). More is $S_6$, less is the difficulty for assessment item (Cheng, 2006).

*H15: Assessment item difficulty level increases as the number of assumptions ($S_7$) available in an item increases.*

In addition to the variables selected from the literature, number of assumptions was also included by the researchers as difficulty increases with the number of implicit and explicit assumptions increases for reaching item solution. Examples of assumptions include some quantitative data, expected process, initial condition/ state and amount of delay.

## 4. Method to Determine the Effectiveness of Chosen Variables

A study is planned to determine the effectiveness of chosen variables to predict difficulty level of assessment items. Difficulty level is dependent variable while independent variables include three task variables, five content variables and seven stimulus variables. Following sub sections present description about item samples, coding of independent variables and tagging of dependent variable.

### 4.1 Sample

Engineering courses can be divided into three categories, courses which focus on application and design, theory courses and courses that emphasize understanding. Thus, each of these categories have at least one dominant difficulty category (task, content and stimulus). Three courses were chosen from three different categories of courses: Digital Systems, Digital Communication, and Design of Algorithms. The selection of items from three different courses ensures the randomization of sample and fairness in the model evaluation for estimating difficulty. The item samples consist of 300 items which were found from assessment instruments, assignments and tutorials designed at premier universities. The chosen items were from Remember (R), Understand (U), Apply (Ap), Analyze (An) and Evaluate (E) cognitive levels (Anderson, Krathwohl, and Bloom, 2001). The specification of item samples used for conducting the survey is as shown in Table 1.

Table 1. Specification of Sample Assessment Items

| Course | R | U | Ap | An | E | Total |
|--------|---|---|----|----|----|-------|
| Design of Algorithms | 8 | 20 | 15 | 42 | 15 | 100 |
| Digital communication | 27 | 48 | 20 | 5 | 0 | 100 |
| Digital systems | 21 | 24 | 45 | 4 | 6 | 100 |
| **Total** | 56 | 92 | 80 | 51 | 21 | 300 |

### 4.2 Independent Variables

Two raters were chosen for coding 300 assessment items according to the chosen independent variables for predicting difficulty level. The raters were education technology researchers working in the area of assessment and evaluation, who were comfortable with the terminology used for variables, the pedagogical theories used and the proposed coding rules. The results of applying coding rule can be subjective to the rater because of the assumptions the rater would have been made. Therefore, the coding for the items was iteratively conducted through discussions among raters with the coding rules discussed discussed in Table 2. Table 3 shows coding rules implemented for a sample item.

### 4.3 Dependent Variables

Six SMEs from premier institutes, two for each course (Digital Systems, Digital Communication, and Design of Algorithms), were chosen on the basis of their respective areas of expertise in teaching the same course. Two SMEs for a course were given 100 items each for tagging the given items for three difficulty levels. The experts were given a detailed written explanation regarding the motivation behind the survey and meaning of low, moderate and high difficulty levels. It is restricted to three levels, since the coding of items may be influenced by subjective assumptions of raters. The correlation was determined among expert judgment about item difficulty level and its value ranged from 0.81 to 0.86 for three courses, which demonstrated a general consensus among experts about difficulty level. Discrepancy of opinion about difficulty level among experts was 10% for moderate and high levels and 6% for low and moderate levels.

Table 2. Coding of Independent Variables

| Independent Variable | Type | Coding | Coding Rule |
|---|---|---|---|
| Number of unknowns ($T_1$) | Numeric | - | Identified from the item stem |
| Number of conditions ($T_2$) | Numeric | - | Identified from the item stem |
| Numerical complexity ($T_3$) | Ordinal | 1 = Simple<br>2 = Moderate<br>3 = Complex | Measured by assigning weights to the numerical values instead of using the numerical values itself. It is assigned three categories: simple, moderate and complex which are coded as 1, 2 and 3 respectively. These categories are decided by SMEs for each course |
| Number of facts ($C_1$) | Numeric | - | Identified from the item solution |
| Number of concepts ($C_2$) | Numeric | - | Identified from the item solution. Concept Effect Table is used for measuring the number of concepts which represents the relationships between the concepts to be learned (Guenel and Asliyan, 2009) |
| Number of procedures ($C_3$) | Numeric | - | Identified from the item solution. Repeated procedure is counted once |
| Combination of knowledge elements ($C_4$) | Ordinal | 1 = F<br>2 = P<br>3 = C<br>4 = F-P<br>5 = F-C<br>6 = C-P<br>7 = F-C-P | Identified from the item stem with Factual (F), Conceptual (C), Procedural (P), Factual-Conceptual (F-C), Factual- Procedural (F-P) and Factual-Conceptual-Procedural (F-C-P) categories. Factual knowledge is considered less difficult as compared to conceptual and procedural knowledge element. Conceptual knowledge is considered to be more difficult as compared to procedural knowledge element (Anderson, Krathwohl, and Bloom, 2001) |
| Number of prerequisite course outcomes from the same course ($C_5$) | Numeric | - | Identified from the item stem. Course outcome map (Nilson, 2009) is used as a tool to measure $C_5$ |
| Item presentation ($S_1$) | Ordinal | 1 = Simple<br>2 = Complex | Identified from the item stem. Coded by SME as simple and complex based on the vocabulary used and the structure of item presentation |
| Number of hints ($S_2$) | Numeric | - | Identified from the item stem |
| Independency of unknowns ($S_3$) | Ordinal | 1 = Not dependent<br>2 = Dependent | Identified from the item stem |
| Usage of technical notations ($S_4$) | Ordinal | 1 = Present<br>2 = Absent | Identified from the item stem |
| Number of inferences ($S_5$) | Numeric | - | Identified from the item stem |
| Number of resources ($S_6$) | Numeric | - | Identified from the item stem |
| Number of assumptions ($S_7$) | Numeric | - | Identified from the item solution |

Table 3. Coding of Independent Variables: An example

Sample Assessment Item: Consider a 4-bit ripple carry adder. Each full-adder is implemented using a 3-input XOR gate, three 2-input AND gates, and one 3-input OR gate.

(a) Draw the circuit diagram for the full-adder. (1 mark)

(b) Consider the delay of each 3-input XOR gate to be 3 nsec, the delay of each 2-input AND gate to be 1 nsec, and the delay of each 3-input OR gate to be 1 nsec. What will be the total time taken by the 4-bit ripple carry adder to perform a successful addition operation? (Hint: Total time is sum of individual full adder propagation delays) (2 marks)

(c) Suppose the two numbers to be added are $(F)_{16}$ and $(1)_{16}$ . Evaluate the sum as a function of time, starting from the beginning of the addition, at every nano-second. (4 marks)

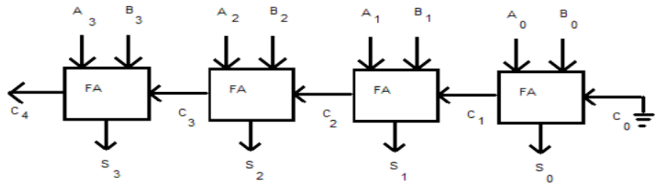

| Independent Variable | Coding | Justification |
|---|---|---|
| Number of unknowns ($T_1$) | 3 | Circuits diagram, Total time, Sum |
| Number of conditions ($T_2$) | 1 | Design of full adder using particular combination of gates |
| Numerical complexity ($T_3$) | 2 | Moderate |
| Number of facts ($C_1$) | 1 | Propagation delay |
| Number of concepts ($C_2$) | 3 | Logic expression, Truth table, Full adder, Ripple carry adder |
| Number of procedures ($C_3$) | 4 | Drawing schematic diagram, Estimating propagation delay, Finding sum of two hexadecimal numbers |
| Combination of knowledge elements ($C_4$) | 7 | Factual-Conceptual-Procedural |
| Number of prerequisite course outcomes from the same course ($C_5$) | 2 | Understand the nature of logic expressions written in terms of logical functions (AND, OR, NOT, NAND, NOR, X-OR, X-NOR), Simplify logical expressions using Karnaugh Maps and Quine- McCluskey Method |
| Item presentation ($S_1$) | 1 | Simple |
| Number of hints ($S_2$) | 1 | Total time is sum of individual full adder propagation delays |
| Independency of unknowns ($S_3$) | 2 | Dependent |
| Usage of technical notations ($S_4$) | 2 | Technical notations absent |
| Number of inferences ($S_5$) | 0 | No inferences |
| Number of resources ($S_6$) | 1 | 4-bit ripple carry adder block diagram |
| Number of assumptions ($S_7$) | 1 | Sum was $(0)_{16}$ before the addition started |

# 5. Data Analysis and Results

This section discusses the model for predicting the difficulty level of assessment items using ordinal regression analysis. 300 items tagged with difficulty level (D) by SMEs and coded with 15 independent variables by raters were used for building the model.

## 5.1 Descriptive Statistics of Independent and Dependent Variables

Table 4 shows mean and standard deviation for numeric variables and frequency table for ordinal variables. The mean number of unknowns for the data set is 2. Certain number of facts, concepts and procedures are needed to attempt each item which is indicated by their mean values. The mean for number of prerequisite course outcomes represents on an average number of course outcomes to be mastered is 3. The mean for number of conditions, resources and inferences denotes majority of the items have these variable value as one. Mostly, all the items dealt with moderate numerical complexity. About 67% of the items had unknowns defined relative to each other. The range for number of conditions and hints is 2, majority of the items have null values for these variables.

**Table 4.** Descriptive Statistics of Dependent and Independent Variables

| Variables | Summary Statistics | |
|---|---|---|
| **Numeric Variables** | **Mean** | **Standard Deviation** |
| $T_1$ | 2.03 | 1.04 |
| $T_2$ | 0.77 | 0.82 |
| $C_1$ | 4.01 | 1.14 |
| $C_2$ | 5.51 | 1.8 |
| $C_3$ | 3.46 | 1.68 |
| $C_5$ | 3.06 | 1.71 |
| $S_2$ | 0.4 | 0.64 |
| $S_5$ | 1 | 1.13 |
| $S_6$ | 0.96 | 0.35 |
| $S_7$ | 0.88 | 0.88 |
| **Ordinal Variables** | **Frequency Table** | |
| $T_3$ | Simple = 28%, Moderate = 49%, Complex = 22% | |
| $C_4$ | C = 4%, CP = 4%, F = 6%, FC = 25%, FCP = 47%, FP = 11%, P = 3% | |
| $S_1$ | Simple = 51%, Complex = 49% | |
| $S_3$ | Dependent = 67%, Not dependent = 33% | |
| $S_4$ | Technical notations absent = 48%, Technical notations present = 52% | |
| D | Low = 27%, Moderate = 35%, High = 38% | |

Table 5 presents the heterogeneous correlations (polyserial (numeric-ordinal) and polychoric (ordinal-ordinal)) among the variables and with assessment item difficulty. *P*-values for correlation tests denote all variables, except $T_3$ ,are correlated well with item difficulty. Correlation matrix indicate a strong relationship between some of the independent variables. Item presentation is correlated with number of inferences and number of resources. Number of concepts is correlated with number of prerequisite course outcomes from the same course, number of unknowns and independency of unknowns. Number of unknowns is correlated with number of facts. Number of procedures is related to number of assumptions and numerical complexity.

**Table 5**. Correlation Matrix of Independent Variables and Difficulty Level

| Variables | T₁ | T₂ | T₃ | C₁ | C₂ | C₃ | C₄ | C₅ | S₁ | S₂ | S₃ | S₄ | S₅ | S₆ | S₇ |
|---|---|---|---|---|---|---|---|---|---|---|---|---|---|---|---|
| **T₁** | 1 | | | | | | | | | | | | | | |
| **T₂** | .24 | 1 | | | | | | | | | | | | | |
| **T₃** | .008 | .03 | 1 | | | | | | | | | | | | |
| **C₁** | .31 | .06 | .22 | 1 | | | | | | | | | | | |
| **C₂** | .74** | .47* | .12 | .36* | 1 | | | | | | | | | | |
| **C₃** | .55** | .18 | .36* | .15 | .39* | 1 | | | | | | | | | |
| **C₄** | .29 | .25 | .39* | .27 | .43* | .66** | 1 | | | | | | | | |
| **C₅** | .2 | .28* | .12 | .4* | .56** | .19 | .34* | 1 | | | | | | | |
| **S₁** | .38* | .18 | .04 | .29* | .22 | .02 | .14 | .44* | 1 | | | | | | |
| **S₂** | .67** | .04 | .01 | .21 | .33* | .52** | .13 | .05 | .17 | 1 | | | | | |
| **S₃** | .71** | .23 | .08 | .57** | .36* | .19 | .22 | .32* | .37* | .3 | 1 | | | | |
| **S₄** | .06 | .09 | .26 | .19 | .16 | .2 | .24 | .19 | .29 | .15 | .06 | 1 | | | |
| **S₅** | .36* | .15 | .04 | .36* | .13 | .06 | .19 | .26 | .72** | .18 | .39* | .57** | 1 | | |
| **S₆** | .19 | .23 | .25* | .26* | .07 | .31* | .17 | .15 | −.59** | .2 | .14 | .21 | .21 | 1 | |
| **S₇** | .56** | .15 | .04 | .1 | .37* | .46* | .27 | .41* | .23 | .24 | .17 | .49* | .41* | -.007 | 1 |
| **D** | .53** | .77** | −.02 | .45** | .65** | .57** | .42* | .47* | .71** | -.47* | .43** | .64** | .52** | -56** | .38** |

*significant at $p < 0.05$, **significant at $p < 0.01$

### 5.2 Ordinal Logistic Regression Model of Item Difficulty Level

Difficulty level of assessment items is ordinal in nature. Ordinal logistic regression is a standard method of analysis in this type of situation which is conducted to develop a model that contains minimum number of independent variables and capable of estimating difficulty level with maximum accuracy. The analysis was started with the null model and search was carried out through models lying in the range between the null and full model using the forward selection algorithm. Criterion based procedure is used to choose significant independent variables. At each stage of this process, independent variable may be added or removed based on Akaike information criterion (*AIC*), Bayesian information criterion (*BIC*) and Deviance. *AIC, BIC* and deviance indicate the ability of the model to fit the data. Deviance indicates the difference between the observed and expected outcomes. *AIC* indicates the ability of the model to represent the data using the chosen set of variables. *BIC* indicates the ability of the model to represent the data using the chosen set of variables and for the chosen size of data set. Less is the value of deviance, better is the goodness of fit for the model. *AIC, BIC* and deviance include penalties for increasing the number of independent variables in the model, and these penalties discourage over fitting. The model with lowest *AIC* and *BIC* values is preferred. Table 6 shows model summary of criterion based procedure regression using the '*clm*' function from the ordinal package (Christensen and Christensen, 2015) in *R* (*R* Core Team, 2015) for selected models.

**Table 6.** Model Summary of Ordinal Logistic Regression

| Model | Independent Variables | *AIC* | *BIC* | Deviance |
|---|---|---|---|---|
| Model 1 | $T_2, T_3, C_1$ | 1029.23 | 1040.82 | 1023.23 |
| Model 2 | $T_2, T_3, C_1, C_2$ | 953.86 | 969.31 | 945.86 |
| Model 3 | $T_2, C_2, C_3, C_5$ | 908.05 | 923.5 | 900.05 |
| Model 4 | $T_2, C_2, C_3, S_1, S_2$ | 885.13 | 904.44 | 875.13 |
| Model 5 | $T_2, C_2, C_3, S_1, S_2, S_4$ | 890.86 | 914.04 | 878.86 |
| Model 6 | $T_2, C_2, C_3, S_1, S_4, S_5$ | 882.46 | 905.64 | 870.46 |
| **Model 7** | $\mathbf{T_2, C_2, C_3, S_1, S_4, S_6}$ | **865.67** | **888.85** | **853.67** |
| Model 8 | $T_2, C_2, C_3, S_2, S_4, S_5, S_6$ | 871.88 | 898.92 | 857.88 |
| Model 9 | $T_2, C_2, C_3, S_2, S_4, S_5, S_6, S_7$ | 873.84 | 904.74 | 857.84 |

**Table 7.** Regression Model Coefficients

| Independent Variables | ML Estimate | Odds Ratio | Std Error | *z* value | $P_r$ (> |z|) |
|---|---|---|---|---|---|
| $T_2$ | 0.28 | 1.32 | 0.07 | 3.58 | 0.01 |
| $C_2$ | 0.54 | 1.71 | 0.15 | 3.6 | 1.54e-12 |
| $C_3$ | 0.47 | 1.59 | 0.07 | 6.71 | 0.05 |
| $S_1$ | 0.51 | 1.66 | 0.16 | 3.18 | 1.94e-06 |
| $S_4$ | 0.23 | 1.25 | 0.07 | 3.28 | 1.97e-07 |
| $S_6$ | -0.48 | 0.61 | 0.16 | -2.96 | 0.001 |

Model - Model 7, Dependent variable - Assessment item difficulty level, $P_r$ - Wald statistics,
Maximum Likelihood(ML) Estimate - $b_1, b_2, b_3, b_4, b_5, b_6$, Intercepts: $a_2 = -5.8$, $a_1 = -3.2$

The results of criterion based ordinal logistic regression shows that Model 7 with number of conditions, number of concepts, number of procedures, item presentation, usage of technical notations and number of resources variables gives the minimum values for *AIC* and *BIC*. Table 7 shows coefficients for the independent variables for Model 7 and their Wald based *p* values. From odds ratio values, it appears that increase in estimates of all the independent variables except number of resources, is associated with higher levels of difficulty level. The predicted probabilities for three difficulty levels is given by equation (1), (2) and (3). Table 8 shows computation of predicted probabilities of difficulty level for sample values of independent variables. The model is validated using test for model fit and multicollinearity.

$$P_{D = High} = 1 / [1 + e^{-(a_2 + b_1 T_2 + b_2 C_2 + b_3 C_3 + b_4 S_1 + b_5 S_4 + b_6 S_6)}] \qquad (1)$$

$$P_{D = Moderate} = 1 / [1 + e^{-(a_1 + b_1 T_2 + b_2 C_2 + b_3 C_3 + b_4 S_1 + b_5 S_4 + b_6 S_6)}] - P_{D = High} \qquad (2)$$

$$P_{D = Low} = 1 - P_{D = Moderate} - P_{D = High} \qquad (3)$$

where $a_i$ - Intercepts, $b_i$ - ML Estimates

**Table 8.** Computation of Difficulty Level for Sample Items

| Item | $T_2$ | $C_2$ | $C_3$ | $S_1$ | $S_4$ | $S_6$ | $P_{D = High}$ | $P_{D = Moderate}$ | $P_{D = Low}$ |
|------|-------|-------|-------|-------|-------|-------|------|------|------|
| Item 1 | 1 | 0 | 1 | 1 | 1 | 2 | 0.005 | 0.05 | 0.93 |
| Item 2 | 1 | 3 | 4 | 1 | 2 | 2 | 0.17 | 0.56 | 0.25 |
| Item 3 | 3 | 6 | 3 | 2 | 2 | 0 | 0.76 | 0.21 | 0.02 |

### 5.2.1. Test for Model Fit

The difference between deviance from reduced model and deviance from full model is used to assess the model fit as shown in Table 9. Model 7 with six independent variables describes the data well since the residual deviance is insignificant. McFadden's Pseudo $R^2$ for this model is estimated as 0.32. Values ranging from $0.2 - 0.4$ indicate excellent model fit. Model fit is also evaluated by estimating the accuracy by which model is able to classify test set assessment items to appropriate difficulty level.

**Table 9.** Results of Drop in Deviance Test

| Model | Likelihood Ratio Statistics | df | $P_r\ ( > Chisq)$ |
|-------|------------------------------|-----|----------------------|
| Reduced model | 189.56 | 6 | < 2.2e-16 |
| Full Model | 155.31 | 15 | < 2.2e-16 |
| Residual | 34.25 | 9 | 0.1 |

**Table 10.** Confusion (Error) Matrix

| | | Digital Systems | | |
|---|---|---|---|---|
| **Accuracy of classification = 82%** | | **Predicted Difficulty Levels** | | |
| | | *Low* | *Moderate* | *High* |
| **Actual Difficulty Levels** | *Low (28)* | **22** | 5 | 1 |
| | *Moderate (32)* | 2 | **24** | 6 |
| | *High (40)* | 0 | 4 | **36** |
| | | **Digital Communication** | | |
| **Accuracy of classification = 79%** | | **Predicted Difficulty Levels** | | |
| | | *Low* | *Moderate* | *High* |
| **Actual Difficulty Levels** | *Low (22)* | **18** | 4 | 0 |
| | *Moderate (37)* | 5 | **29** | 3 |
| | *High (41)* | 1 | 8 | **32** |
| | | **Design of Algorithms** | | |
| **Accuracy of classification = 80%** | | **Predicted Difficulty Levels** | | |
| | | *Low* | *Moderate* | *High* |
| **Actual Difficulty Levels** | *Low (30)* | **24** | 6 | 0 |
| | *Moderate (40)* | 4 | **31** | 5 |
| | *High (30)* | 0 | 5 | **25** |

Equations (1), (2) and (3) are used to predict difficulty levels for the sample assessment items for three courses as shown in Table 10. All correct guesses are located in the diagonal of the table. Average accuracy of classification for Digital Systems, Digital Communication, and Design of Algorithms courses is 0.81, a value that is acceptable.

*5.2.2 Test for Multicollinearity*

Two or more independent variables that are correlated, explain almost the same variability in the dependent variable. Test for multicollinearity estimates variance inflation factor (*VIF*) which quantifies how much of the variance of the estimated coefficients are inflated. *VIF* for all the independent variables in the model is shown in Table 11 which indicates no multicollinearity in the chosen model as all independent variables have VIF values less than recommended maximum *VIF* value of 5.

**Table 11.** Results of Multicollinearity Test

| Independent Variables | $T_2$ | $C_2$ | $C_3$ | $S_1$ | $S_4$ | $S_6$ |
|---|---|---|---|---|---|---|
| Variance Inflation Factor | 2.4 | 3.68 | 1.89 | 4.42 | 3.53 | 1.67 |

# 6. Discussion and Conclusion

Understanding of engineering concepts in depth is a pre-requisite for applying engineering knowledge and skills in practical situations. Formative and summative assessments designed using assessment items of a wide range of difficulty level, will facilitate instructors to assess the mastery of course outcomes. This paper presented a model for estimating difficulty level as perceived by experts.

The learner independent and generic variables to estimate the difficulty level were identified as number of unknowns, number of conditions, numerical complexity, number of facts, number of concepts, number of procedures, combination of knowledge elements, number of prerequisite course outcomes from the same course, item presentation, number of hints, independency of unknowns, usage of technical notations, number of inferences, number of resources and number of assumptions. All the variables, except numerical complexity, were found correlated with difficulty level. Difficulty level increases with difficulty of procedures and not with the numbers in the procedure.

A model is proposed for estimating difficulty level of a given assessment item using six variables. The suitability of the variables (number of conditions (H2), number of concepts (H5), number of procedures (H6), item presentation (H9), usage of technical notations (H12) and number of resources (H14)) chosen for the model complement previous studies in this area. Difficulty level is primarily determined by the manner in which the item is presented (stimuli) to the students.

As future work, the reliability of the model is to be verified on a bigger data set. While the proposed model estimates the difficulty level of assessment items, the same model can be used to design a set of guidelines to instructors for designing assessment items of a specified difficulty level.

References

Ahmed, A., & Pollitt, A. (1999). Curriculum demands and question difficulty. In *IAEA conference, Bled, Slovenia*.

Anderson, L. W., Krathwohl, D. R., & Bloom, B. S. (2001). *A taxonomy for learning, teaching, and assessing: A revision of Bloom's taxonomy of educational objectives*. Allyn & Bacon.

Aryadoust, V. (2013, April). Predicting item difficulty in a language test with an Adaptive Neuro Fuzzy Inference System. In *2013 IEEE Workshop on Hybrid Intelligent Models and Applications (HIMA)* (pp. 43-50). IEEE.

Aryadoust, V., & Goh, C. C. M. (2014). Predicting listening item difficulty with language complexity measures: A comparative data mining study. CaMLA Working Papers, 2014-01. Ann Arbor, MI: CaMLA.

Aryadoust, V., & Baghaei, P. (2016). Does EFL readers' lexical and grammatical knowledge predict their reading ability? Insights from a perceptron artificial neural network study. Educational Assessment, 21(2), 135-156. DOI: 10.1080/10627197.2016.1166343.

Banerjee, S., Rao, N. J., & Ramanathan, C. (2015, October). Rubrics for assessment item difficulty in engineering courses. In *Frontiers in Education Conference (FIE), 2015. 32614 2015. IEEE* (pp. 1-8). IEEE.

Buck, G., & Tatsuoka, K. (1998). Application of the rule-space procedure to language testing: Examining attributes of a free response listening test. Language Testing, 15(2), 119–157. doi: 10.1177/026553229801500201

Buck, G., Tatsuoka, K., & Kostin, I. (1997). The subskills of reading: Rule-space analysis of a multiple-choice test of second language reading comprehension. Language Learning, 47(3), 423–466. doi: 10.1111/0023-8333.00016

Cheng, L. S. (2006, May). On varying the difficulty of test items. In *Annual Conference of the International Association for Educational Assessment*.

Chon, Y. V., & Shin, T. (2010). Item difficulty predictors of a multiple-choice reading test. *English Teaching*, 65(4), 257-282.

Christensen, R. H. B., & Christensen, M. R. H. B. (2015). Package 'ordinal'.

Dhillon, D. (2003). Predictive Models of Question Difficulty–A Critical Review of the Literature.

Embretson, S. E., & Daniel, R. C. (2008). Understanding and quantifying cognitive complexity level in mathematical problem solving items. *Psychology Science*, 50(3), 328.

Fisher-Hoch, H., & Hughes, S. (1996). What makes mathematics exam questions difficult. *British Educational Research Association, University of Lancaster, England*.

Fisher-Hoch, H., Hughes, S., & Bramley, T. (1994). What makes GCSE examination questions difficult? Outcomes of manipulating difficulty of GCSE questions.

Freedle, R., & Kostin, I. (1991). The prediction of SAT reading comprehension item difficulty for expository prose passages (ETS Research Report No. RR-91-29). Princeton, NJ: Educational Testing Service.

Freedle, R., & Kostin, I. (1999). Does the text matter in a multiple-choice test of comprehension?

Gan, Z. (2011). Second language task difficulty: Reflections on the current psycholinguistic models. *Theory and Practice in Language Studies*, 1(8), 921-927.

Günel, K., & Asliyan, R. (2009). Determining difficulty of questions in intelligent tutoring systems. *TOJET: The Turkish Online Journal of Educational Technology*, 8(3).

Hatzilygeroudis, I., Koutsojannis, C., Papavlasopoulos, C., & Prentzas, J. (2006, July). Knowledge-based adaptive assessment in a Web-based intelligent educational system. In *Sixth IEEE International Conference on Advanced Learning Technologies (ICALT'06)* (pp. 651-655). IEEE.

Hughes, S., Pollitt, A., & Ahmed, A. (1998). The development of a tool for gauging the demands of GCSE and A Level exam questions. *BERA, Queen's University Belfast*.

Jiu, K. A., & Park, C. (2004). The Prediction of English Item Difficulty in College Scholastic Ability Test.

Karahoca, A., & Karahoca, D. (2009, September). ANFIS supported question classification in computer adaptive testing (CAT). In *Soft Computing, Computing with Words and Perceptions in System Analysis, Decision and Control, 2009. ICSCCW 2009. Fifth International Conference on* (pp. 1-4). IEEE.

Katz, I. R., Lipps, A. W., & Trafton, J. G. (2002). Factors affecting difficulty in the generating examples item type. *ETS Research Report Series*, 2002(1), i-40.

Khan, J. I., Hardas, M., & Ma, Y. (2005, September). A study of problem difficulty evaluation for semantic network ontology based intelligent courseware sharing. In *Proceedings of the 2005 IEEE/WIC/ACM International Conference on Web Intelligence* (pp. 426-429). IEEE Computer Society.

Klemola, T. (2000). A cognitive model for complexity metrics. In *4th International ECOOP Workshop on Quantitative Approaches in Object-Oriented Software Engineering (ECOOP 2000), Sophia Antipolis*.

Koutsojannis, C., Beligiannis, G., Hatzilygeroudis, I., Papavlasopoulos, C., & Prentzas, J. (2007). Using a hybrid AI approach for exercise difficulty level adaptation. *International Journal of Continuing Engineering Education and Life Long Learning*, 17(4-5), 256-272.

Kunichika, H., Urushima, M., Hirashima, T., & Takeuchi, A. (2002, December). A computational method of complexity of questions on contents of english sentences and its evaluation. In *Computers in Education, 2002. Proceedings. International Conference on* (pp. 97-101). IEEE.

Kuo, R., Lien, W. P., Chang, M., & Heh, J. S. (2004). Analyzing problem's difficulty based on neural networks and knowledge map. *Educational Technology & Society*, *7*(2), 42-50.

Lee, F. L., & Heyworth, R. (2000). Problem complexity: A measure of problem difficulty in algebra by using computer. *EDUCATION JOURNAL-HONG KONG-CHINESE UNIVERSITY OF HONG KONG-*, *28*(1), 85-108.

Li, T., & Sambasivam, S. E. (2003). Question Difficulty Assessment in Intelligent Tutor Systems for Computer Architecture. *Information Systems Education Journal*, *1*(51).

Liu, J., Sha, S., Zheng, Q., & Chen, L. (2010, November). Ranking Difficulty of Knowledge Units Based on Learning Dependency. In *e-Business Engineering (ICEBE), 2010 IEEE 7th International Conference on* (pp. 77-82). IEEE.

Lumley, T., Routitsky, A., Mendelovits, J., & Ramalingam, D. (2012). A framework for predicting item difficulty in reading tests.

Matters, G. (2010). A three-way classification of sources of item difficulty in tests and examinations.

Moskal, B. M. (2000). Scoring Rubrics: How?.

Nakamura, Y., Kuwabara, T., & Takeda, K. (1998, July). Factors determining the difficulty of practice problems in a computer language textbook. In *Computer Human Interaction, 1998. Proceedings. 3rd Asia Pacific* (pp. 422-428). IEEE.

Nilson, L. B. (2009). *The graphic syllabus and the outcomes map: Communicating your course* (Vol. 137). John Wiley & Sons.

Nosike, M. C., & Akukwe, A. C. (2013). Task Number and Cognitive Complexity as Determinants of Difficulty Levels In Multiple–Choice Test Items. *West African Journal of Industrial and Academic Research*, *8*(1), 227-235.

Osterlind, S. J., & Friedman, S. J. (1999). Constructing test items: Multiple-choice, constructed response, performance, and other formats. *Journal of Educational Measurement*, *36*(3), 267-270.

Ozuru, Y., Rowe, M., O'Reilly, T., & McNamara, D. S. (2008). Where's the difficulty in standardized reading tests: The passage or the question?. *Behavior Research Methods*, *40*(4), 1001-1015.

Perkins, K., Gupta, L., & Tammana, R. (1995). Predicting item difficulty in a reading comprehension test with an artificial neural network. *Language testing*, *12*(1), 34-53.

Pollitt, A., & Ahmed, A. (1999, May). A new model of the question answering process. In *IAEA conference, Slovenia, May 1999*.

Schmeiser, C. B., & Whitney, D. R. (1973). The Effect of Selected Poor Item-Writing Practices on Test Difficulty, Reliability and Validity: A Replication.

Sung, P. J., Lin, S. W., & Hung, P. H. (2015). Factors Affecting Item Difficulty in English Listening Comprehension Tests. *Universal Journal of Educational Research*, *3*(7), 451-459.

The case for the construct validity of TOEFL's minitalks. Language Testing, 16(1), 2–32. doi: 10.1177/026553229901600102

Turner, R., & Adams, R. J. (2012). Some drivers of test item difficulty in mathematics: an analysis of the competency rubric.

Vincenti, W. G. (1990). What Engineers Know and How They Know it Analytical Studies from Aeronautical History.

**Author's Bioprofile:**

1. Shilpi Banerjee, International Institute of Information Technology, Bangalore, India
shilpibanerjee1@gmail.com, 09731128466
Shilpi Banerjee works as Assistant Professor at Azim Premji University (APU), in the School of Continuing Education. Shilpi has a Bachelor's degree from Nagpur University, MTech from National Institute of Technology, Bhopal and PhD in Educational Assessment from International Institute of Information Technology (IIIT), Bangalore. At APU, she is involved in the design and review of large scale tests used for assessing elementary grades students learning outcomes and in analyzing students' test results to understand how well they have attained these outcomes. She has conducted several training workshops for engineering college faculties and school teachers on learning outcomes, assessment, course design, item bank creation and tutoring. She has over 10 publications in reputed international conferences and journals.

2. N.J.Rao, Indian Institute of Science, Bangalore, India
njraoiisc@gmail.com, 09343537734
N.J. Rao was the Chairman of CEDT (Centre for Electronics Design and Technology, IISc during 1981 – 1996, and Chairman, Department of Management Studies during 1998 – 2006, and superannuated as Professor at CEDT in July 2006. He is presently a Consulting Professor at International Institute of Information Technology (IIIT), Bangalore, a member of several committees associated with NBA, and a member of the Core Committee that defined the new Accreditation processes of NAAC. His research areas included Control Systems and System Dynamics. His present research interests include higher education, pedagogy and education technologies.